\documentclass{article}

\usepackage{arxiv}
\usepackage[utf8]{inputenc}
\usepackage[T1]{fontenc}
\usepackage{hyperref}
\usepackage{url}
\usepackage{amsmath,amssymb}
\usepackage{graphicx}
\usepackage{booktabs}
\usepackage{natbib}
\usepackage{xcolor}
\usepackage{soul}
\usepackage{tikz}
\usepackage{tabularx}
\usepackage{booktabs}
\usepackage{makecell}
\usepackage{ragged2e}
\newcolumntype{Y}{>{\RaggedRight\arraybackslash}X} % For wrapped text
\usetikzlibrary{shapes.geometric, arrows.meta, positioning, calc, shadows, backgrounds, fit}

\title{Deep Learning for Contextualized NetFlow-Based Network Intrusion Detection: Methods, Data, Evaluation and Deployment}

\author{\textbf{Abdelkader El Mahdaouy}\\
College of Computing\\
Mohammed VI Polytechnic University\\
Ben Guerir, Morocco\\
\texttt{firstname.lastname@um6p.ma}
\And \textbf{Issam Ait Yahia}\\
College of Computing\\
Mohammed VI Polytechnic University\\
Ben Guerir, Morocco\\
\texttt{firstname.lastname@um6p.ma}
\And 
\textbf{Soufiane Oualil}\\
Hassan II University\\
Casablanca, Morocco\\
\texttt{soufiane.oualil-etu@etu.univh2c.ma}
\And
\textbf{Ismail Berrada}\\
College of Computing\\
Mohammed VI Polytechnic University\\
Ben Guerir, Morocco\\
\texttt{firstname.lastname@um6p.ma}
}

\begin{document}

\maketitle

% -------------------------------------------------------------
% ABSTRACT
% -------------------------------------------------------------
\begin{abstract}
Network Intrusion Detection Systems (NIDS) have progressively shifted from signature-based techniques toward machine learning and, more recently, deep learning methods. Meanwhile, the widespread adoption of encryption has reduced payload visibility, weakening inspection pipelines that depend on plaintext content and increasing reliance on flow-level telemetry such as NetFlow and IPFIX. Many current learning-based detectors still frame intrusion detection as per-flow classification, implicitly treating each flow record as an independent sample. This assumption is often violated in realistic attack campaigns, where evidence is distributed across multiple flows and hosts, spanning minutes to days through staged execution, beaconing, lateral movement, and exfiltration. This paper synthesizes recent research on context-aware deep learning for flow-based intrusion detection. We organize existing methods into a four-dimensional taxonomy covering temporal context, graph or relational context, multimodal context, and multi-resolution context. Beyond modeling, we emphasize rigorous evaluation and operational realism. We review common failure modes that can inflate reported results, including temporal leakage, data splitting, dataset design flaws, limited dataset diversity, and weak cross-dataset generalization. We also analyze practical constraints that shape deployability, such as streaming state management, memory growth, latency budgets, and model compression choices. Overall, the literature suggests that context can meaningfully improve detection when attacks induce measurable temporal or relational structure, but the magnitude and reliability of these gains depend strongly on rigorous, causal evaluation and on datasets that capture realistic diversity.
\end{abstract}

% -------------------------------------------------------------
% INTRODUCTION
% -------------------------------------------------------------
\section{Introduction}
\label{sec:intro}

Network Intrusion Detection Systems (NIDS) have traditionally relied on signature-based or rule-based mechanisms to identify known threats. Over the past decade, this paradigm has progressively shifted toward Machine Learning (ML) and Deep Learning (DL), driven by the need to detect zero-day attacks, polymorphic malware, and adaptive adversarial behavior. Learning-based NIDS enable anomaly detection even in the absence of explicit signatures, by modeling deviations from learned behavioral patterns~\citep{shone2018,sharma2024encrypted,IdrissiABB23,abdelhalim2025,farhan2025}.

In operational environments, however, network intrusion detection increasingly operates under constraints that limit deep packet inspection, including privacy regulations, high-throughput requirements, and widespread encryption. These factors have accelerated the adoption of \emph{flow-based} telemetry, where traffic is summarized into compact records exported by NetFlow or IP Flow Information Export (IPFIX)~\citep{rfc3954,rfc7011ipfix,hofstede2014flow,sperotto2010overview}. Flow records aggregate communications using a five-tuple key and associated statistics, enabling scalable monitoring while avoiding payload inspection. Consequently, modern NIDS infer malicious behavior from statistical and behavioral features rather than raw content, trading semantic richness for scalability and privacy~\citep{umer2017,farhan2025,IdrissiAMBYB25}.

A central methodological question in flow-based NIDS is whether individual flows can be treated as independent samples. In practice, many threats are inherently \emph{contextual}: they manifest through dependencies across flows over time (e.g., beaconing and staged exfiltration), across topology (e.g., coordinated scanning and lateral movement), and across heterogeneous signals (e.g., DNS activity and anomalous TLS handshakes). Treating each flow as an IID instance discards precisely the relational and sequential structure that characterizes multi-stage and stealthy attacks~\citep{sommer2010outside,deng2023,farhan2025}. Moreover, high per-flow accuracy is insufficient in operational settings: due to class imbalance and the base-rate fallacy, even detectors with low false-positive rates can overwhelm analysts with alerts~\citep{axelsson2000base}. These observations motivate \emph{context-aware} learning, where models explicitly represent dependencies across flows.

Most existing NIDS literature nevertheless frames intrusion detection as a per-flow classification or anomaly detection problem, using feature vectors extracted from individual flow records and training classifiers such as support vector machines, random forests, or neural networks~\citep{zhang2019,pektas2018,su2020,IdrissiAMBYB25,abdelhalim2025}. While this paradigm benefits from simplicity and accessible datasets, multiple studies have documented fundamental limitations.

\begin{itemize}
    \item \textbf{Representational limitations.} Per-flow models ignore sequential and relational dependencies, limiting their ability to detect coordinated or multi-stage attacks. \citet{sommer2010outside} argued that generic intrusion detection is fundamentally constrained by the semantic gap between observable features and malicious behavior. This issue is amplified by the base-rate fallacy, which shows that even highly accurate detectors generate overwhelming false positives in realistic settings~\citep{axelsson2000base}. These effects persist in modern deep learning-based NIDS~\citep{abdelhalim2025,farhan2025}.
    
    \item \textbf{Evaluation artifacts.} Reported performance on benchmark datasets is often inflated by methodological flaws, including feature normalization across training and test sets, random data splitting, and unrealistic attack segmentation~\citep{zavrak2020,arp2022dos,apruzzese2022crosseval,koukoulis2025self,bad_smells}. When evaluated using time-ordered splits, cross-dataset testing, or realistic base rates, performance degrades substantially~\citep{arp2022dos,apruzzese2022crosseval,LAYEGHY2023108692,koukoulis2025self}.
    
    \item \textbf{Benchmark dataset limitations.} Widely used datasets such as CIC-IDS-2017 and CSE-CIC-IDS-2018 exhibit systematic issues, including labeling errors, insufficient documentation of collection procedures, and artificial temporal separation between benign and malicious traffic~\citep{Engelen20217,lanvin2022errors}. More broadly, \citet{bad_smells} identify recurring design flaws in NIDS benchmarks, including limited feature diversity, ambiguous ground truth, distribution collapse, and synthetic diversity generation. These issues correlate with poor generalization in operational environments. Recent analyses further show that dataset diversity, measured using metrics such as the Vendi score and Jensen--Shannon divergence, is a primary determinant of model transferability~\citep{nougnanke2025diversity,friedman2023vendi,goldschmidt2025survey}.
\end{itemize}

To address these limitations, recent research has moved beyond per-flow classification toward \emph{context-aware deep learning} for flow-based NIDS. Broadly, context is captured through four complementary dimensions: (i) \textbf{temporal modeling} of sequences and sessions, (ii) \textbf{graph-based modeling} of communication structure and host relationships, (iii) \textbf{multimodal fusion} of heterogeneous signals available at scale (flows, DNS, TLS metadata, alerts), and (iv) \textbf{multi-resolution aggregation} that combines local and global views of activity. These approaches aim to detect threats whose evidence is distributed across time and topology while remaining compatible with flow-level telemetry.

Progress in context-aware NIDS is limited by current evaluation practices. Despite well-known dataset issues, many studies still rely on protocols that break temporal order, allow feature leakage, and ignore cross-dataset validation~\citep{engelen2021troubleshooting,apruzzese2022crosseval,cantone2024crossdataset}. As a result, models often achieve high benchmark scores while failing under realistic deployment conditions. Credible advances therefore require evaluation frameworks that enforce strict temporal causality, explicitly prevent information leakage, and mandate validation across heterogeneous network environments~\citep{bad_smells,goldschmidt2025survey}.

\paragraph{Contributions and scope.}
This paper surveys recent advances in \emph{context-aware deep learning for flow-based NIDS}. We make four contributions:
\begin{enumerate}
    \item We propose a \textbf{four-dimensional taxonomy} that categorizes methods by the type of context they model: temporal, graph, multimodal, and multi-resolution.
    \item We review \textbf{architectural approaches} to context modeling, including sequential models, transformers, temporal convolutional networks, graph neural networks, heterogeneous graphs, self-supervised learning, and multimodal fusion, with emphasis on scalability and streaming feasibility.
    \item We provide a \textbf{methodological analysis} of evaluation protocols and dataset quality, synthesizing recent work on dataset flaws, diversity, and generalization failures~\citep{engelen2021troubleshooting,lanvin2022errors,liu2022errorprevalence,bad_smells,nougnanke2025diversity,apruzzese2022crosseval,IdrissiAMBYB25,goldschmidt2025survey}.
    \item We discuss \textbf{deployment considerations}, including state management, resource constraints, robustness to concept drift, and cross-dataset validation, and derive evidence-based guidelines for practitioners.
\end{enumerate}

Beyond these four dimensions, emerging research directions include unsupervised and self-supervised learning for temporal and relational modeling~\citep{ruids2023}, privacy-preserving intrusion detection using cryptographic techniques~\citep{primia2024}, and explainable AI for security analytics~\citep{sobchuk2024sequential,xaiids2023,AL2025110145,SHARMA2024121751,JAVEED2024103540}. These trends highlight a broader shift toward NIDS architectures that are not only accurate, but also robust, privacy-aware, and operationally viable.

\paragraph{Structure.}
The remainder of this paper is organized as follows. Section~\ref{sec:taxonomy} presents the proposed taxonomy and discusses evaluation rigor. Sections~\ref{sec:temporal} and~\ref{sec:graph} analyze temporal and graph-based modeling, respectively. Section~\ref{sec:multimodal} covers multimodal and multi-resolution methods. Section~\ref{sec:evaluation} reviews datasets and evaluation protocols, while Section~\ref{sec:implementation} addresses implementation and deployment challenges. We conclude in Section~\ref{sec:conclusion}.

% -------------------------------------------------------------
% TAXONOMY AND EVALUATION RIGOR
% -------------------------------------------------------------
\section{Four-Dimensional Taxonomy}
\label{sec:taxonomy}

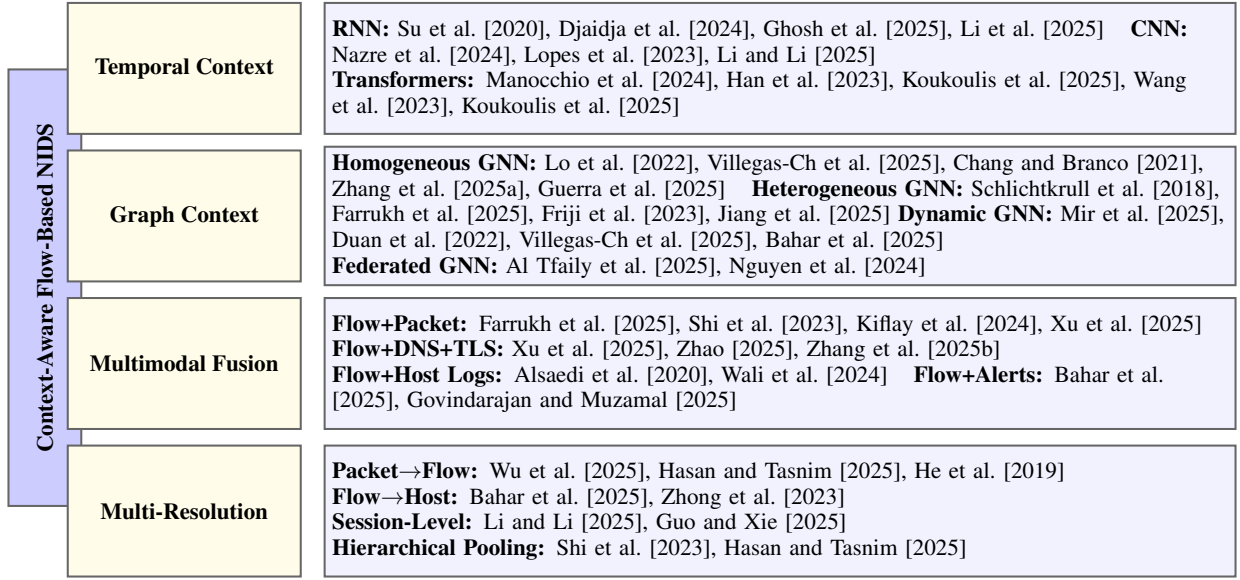
\begin{figure*}[t]
\centering
\resizebox{\textwidth}{!}{
\begin{tikzpicture}[node distance=0.1cm, font=\footnotesize]
  % Sidebar
  \node[rectangle, draw=black!60, line width=0.8pt, fill=blue!20,
        minimum width=6cm, minimum height=1cm, align=center, rotate=90] (sidebar) {\textbf{Context-Aware Flow-Based NIDS}};
  
  % Temporal row
  \node[rectangle, draw=black!60, line width=0.8pt, fill=yellow!10,
        minimum width=3.2cm, minimum height=1.8cm, align=center, right=0.3cm of sidebar] (tempcat) {\textbf{Temporal Context}};
  \node[rectangle, draw=black!60, line width=0.8pt, fill=blue!5,
        minimum width=12.5cm, minimum height=1.8cm, align=left, right=0.3cm of tempcat,
        text width=12.3cm, inner sep=3pt] (tempcontent) {
    \textbf{RNN:} \citet{su2020}, \citet{djaidja2024}, \citet{ghosh2025tacnet}, \citet{adfcnn2025}\hspace{0.8em}
    \textbf{CNN:} \citet{nazre2024tcn}, \citet{lopes2023network}, \citet{li2025lightweight}\\
    \textbf{Transformers:} \citet{manocchio2023flowtransformer}, \citet{han2023network}, \citet{koukoulis2025self}, \citet{ruids2023}, \citet{koukoulis2025self}
  };
  
  % Graph row
  \node[rectangle, draw=black!60, line width=0.8pt, fill=yellow!10,
        minimum width=3.2cm, minimum height=1.8cm, align=center, below=0.2cm of tempcat] (graphcat) {\textbf{Graph Context}};
  \node[rectangle, draw=black!60, line width=0.8pt, fill=blue!5,
        minimum width=12.5cm, minimum height=1.8cm, align=left, right=0.3cm of graphcat,
        text width=12.3cm, inner sep=3pt] (graphcontent) {
    \textbf{Homogeneous GNN:} \citet{lo2022egraphsage}, \citet{villegas2025dynamic}, \citet{chang2021modified}, \citet{zhang2025resacag}, \citet{guerra2025graphids}\hspace{0.8em}
    \textbf{Heterogeneous GNN:} \citet{schlichtkrull2018rgcn}, \citet{farrukh2024xgnid}, \citet{efficientgnn2023}, \citet{jiang2025hgnn}
    \textbf{Dynamic GNN:} \citet{mir2025dynkdd}, \citet{duan2023dynamic}, \citet{villegas2025dynamic}, \citet{messai2025continuum}\hspace{0.8em}\\
    \textbf{Federated GNN:} \citet{Tfaily2025fedgatsage}, \citet{nguyen2024fedmse}
  };
  
  % Multimodal row
  \node[rectangle, draw=black!60, line width=0.8pt, fill=yellow!10,
        minimum width=3.2cm, minimum height=1.8cm, align=center, below=0.2cm of graphcat] (mmcat) {\textbf{Multimodal Fusion}};
  \node[rectangle, draw=black!60, line width=0.8pt, fill=blue!5,
        minimum width=12.5cm, minimum height=1.8cm, align=left, right=0.3cm of mmcat,
        text width=12.3cm, inner sep=3pt] (mmcontent) {
    \textbf{Flow+Packet:} \citet{farrukh2024xgnid}, \citet{shi2023mhpnnids}, \citet{kiflay2024multimodal}, \citet{xu2025fewshot}\hspace{0.8em}
    \textbf{Flow+DNS+TLS:} \citet{xu2025fewshot}, \citet{zha2025dmids}, \citet{zhang2025norns}\\
    \textbf{Flow+Host Logs:} \citet{toniots}, \citet{wali2023meta}\hspace{0.8em}
    \textbf{Flow+Alerts:} \citet{messai2025continuum}, \citet{govindarajan2025}
  };
  
  % Multi-resolution row
  \node[rectangle, draw=black!60, line width=0.8pt, fill=yellow!10,
        minimum width=3.2cm, minimum height=1.8cm, align=center, below=0.2cm of mmcat] (mrcat) {\textbf{Multi-Resolution}};
  \node[rectangle, draw=black!60, line width=0.8pt, fill=blue!5,
        minimum width=12.5cm, minimum height=1.8cm, align=left, right=0.3cm of mrcat,
        text width=12.3cm, inner sep=3pt] (mrcontent) {
    \textbf{Packet$\rightarrow$Flow:} \citet{wu2025uninet}, \citet{hasan2025iot}, \citet{he2019novel}\hspace{0.8em}\\
    \textbf{Flow$\rightarrow$Host:} \citet{messai2025continuum}, \citet{zhong2023dynamic}\\
    \textbf{Session-Level:} \citet{li2025lightweight}, \citet{guo2025research}\hspace{0.8em}\\
    \textbf{Hierarchical Pooling:} \citet{shi2023mhpnnids}, \citet{hasan2025iot}
  };
\end{tikzpicture}
}%
\caption{Taxonomy of context-aware deep learning for flow-based network intrusion detection. The taxonomy organizes methods by the type of context they exploit: temporal, graph, multimodal, and multi-resolution. }
\label{fig:taxonomy}
\end{figure*}
We organize context-aware flow-based NIDS methods along four complementary dimensions, each corresponding to a distinct form of dependency exploited by the detection model. As illustrated in Figure~\ref{fig:taxonomy}, these dimensions reflect how contextual information is represented and integrated into the learning process.

\begin{itemize}
    \item \textbf{Temporal context} captures sequential dependencies across flows over time, enabling detection of attacks whose evidence is distributed across multiple events. Temporal models reason over ordered sequences within sliding windows or sessions, making them suitable for identifying multi-stage and low-and-slow attacks. Correct evaluation in this setting requires strict enforcement of temporal causality to prevent information leakage and ensure realistic performance estimates~\citep{corsini2021evaluation}.

    \item \textbf{Graph context} models relational dependencies among network entities such as hosts, services, flows, or packets. By representing traffic as communication graphs, graph-based methods capture structural patterns and coordinated behavior that are not observable at the individual flow level. This paradigm is particularly effective for detecting distributed and lateral attacks, but introduces challenges related to dynamic graph construction, scalability, and privacy in operational environments~\citep{bilot2023survey,zhong2024survey}.

    \item \textbf{Multimodal fusion} integrates heterogeneous sources of information, including flow records, DNS activity, TLS metadata, alerts, and host-level logs~\citep{shi2023mhpnnids,kiflay2024multimodal}. Combining complementary modalities provides richer semantic context and improves detection coverage, but requires careful alignment, normalization, and handling of modality imbalance.

    \item \textbf{Multi-resolution aggregation} captures activity at different levels of granularity, ranging from packets and flows to sessions and hosts~\citep{wu2025uninet, hasan2025iot}. This dimension enables models to jointly detect fine-grained stealthy behavior and coarse-grained volumetric anomalies, improving robustness across diverse attack types.
\end{itemize}

Together, these four dimensions provide a unifying framework for analyzing context-aware NIDS architectures. Individual systems may exploit one or more dimensions simultaneously, but their effectiveness ultimately depends on how well contextual dependencies are represented, integrated, and evaluated under realistic deployment constraints.

\section{Temporal Modeling}
\label{sec:temporal}

Temporal modeling addresses the limitations of per-flow intrusion detection by explicitly capturing dependencies across chronologically ordered network flows. Many real-world attacks manifest through sequences of related events rather than isolated flows, including low-and-slow exfiltration, reconnaissance-to-exploitation pipelines, and coordinated multi-stage intrusions. Temporal approaches therefore enable detectors to reason over evolving traffic behavior instead of static flow attributes.

The effectiveness of temporal modeling depends not only on architectural choices but also on how sequences are constructed and evaluated. Prior work shows that improper temporal alignment, global normalization, or random data partitioning can introduce temporal leakage, leading to overly optimistic performance estimates that do not generalize to deployment settings~\citep{corsini2021evaluation}. This section reviews major temporal modeling paradigms for flow-based NIDS and discusses their methodological assumptions and practical limitations. Table~\ref{tab:temporal_char} summarizes the main classes of temporal context models used in flow-based NIDS, highlighting their architectural mechanisms, the types of attack dynamics they capture, and the evaluation constraints.

\begin{table*}[t]
\centering
\footnotesize
\setlength{\tabcolsep}{3.5pt} % Reduced from 4pt for tighter spacing
\caption{Characteristics of Temporal Context Methods}
\label{tab:temporal_char}
\begin{tabularx}{\textwidth}{@{}
    >{\raggedright\arraybackslash}p{1.65cm}
    >{\raggedright\arraybackslash}X
    >{\raggedright\arraybackslash}X
    >{\raggedright\arraybackslash}X
    >{\raggedright\arraybackslash}X
    >{\raggedright\arraybackslash}p{2.25cm}
@{}}
\toprule
\textbf{Method} & \textbf{Key Mechanisms} & \textbf{Attack Profile} & \textbf{Key Limitations / Constraints} & \textbf{Critical Evaluation Requirements} & \textbf{Works} \\
\midrule
RNNs (LSTM/GRU) &
Stateful hidden vectors; unidirectional (causal) vs.\ bidirectional variants. &
Multi-stage attacks, low-and-slow exfiltration, irregular beaconing. &
Performance degrades on long-range dependencies under truncated BPTT; sequential inference limits throughput. &
Bidirectional models use future context and are offline; online deployment requires unidirectional or chunked streaming. &
\citet{su2020,djaidja2024,ghosh2025tacnet,adfcnn2025} \\
\midrule
Temporal CNNs (TCN) &
Dilated causal convolutions over fixed windows; residual blocks. &
Short-range anomalies, burst attacks, protocol state machines. &
Sequence modeling limited by receptive field and window segmentation. &
Causal padding; lagged statistics only; chronological splits. &
\citet{nazre2024tcn,lopes2023network,li2025lightweight} \\
\midrule
Transformers &
Self-attention with positional encoding and causal masking. &
Long-range campaigns, delayed stages, distributed beaconing. &
Quadratic cost with sequence length; highly sensitive to leakage. &
Strict causal attention; no future-derived features; time-ordered splits. &
\citet{manocchio2023flowtransformer,han2023network,koukoulis2025self,ruids2023} \\
\midrule
Self-Supervised Temporal &
Masked reconstruction or contrastive learning on sequences. &
Zero-day and novel temporal patterns. &
Augmentation design sensitive; requires large unlabeled data. &
Augmentations must respect temporal order; cross-dataset validation. &
\citet{ruids2023,koukoulis2025self} \\
\bottomrule
\end{tabularx}
\end{table*}

\subsection{Recurrent Neural Networks}

Recurrent neural networks model flow sequences incrementally by maintaining hidden states that summarize past observations. Unidirectional variants preserve causality and are suitable for online detection, whereas bidirectional models exploit future context and are therefore limited to offline analysis. Long Short-Term Memory (LSTM) networks have demonstrated improved detection of temporally structured attacks on datasets such as CIC-IDS-2017 and CTU-13 when evaluated under leakage-free protocols that respect temporal ordering~\citep{corsini2021evaluation,su2020,djaidja2024}. 

Despite their effectiveness, RNN-based models suffer from well-known limitations, including vanishing gradients on long sequences, high training and inference costs, and sensitivity to overfitting in heterogeneous or multi-resolution traffic scenarios~\citep{farrukh2024xgnid,zhang2019}. These constraints limit their scalability in high-throughput or real-time environments. ~\citep{ghosh2025tacnet} proposed TACNet, a novel framework combining multi-scale CNNs with LSTM and temporal attention mechanisms for IoT/IIoT intrusion detection, achieving nearly 100\% accuracy across multiple datasets. ~\citep{adfcnn2025} introduced ADFCNN-BiLSTM, which uses deformable convolution and multi-head attention to adaptively extract spatial features while BiLSTM mines temporal dependencies.

\subsection{Spatio-Temporal Feature Learning with TCNs and Hybrid Models}

Although Temporal Convolutional Networks (TCNs) are often described as temporal models, most TCN-based approaches in flow-based NIDS do not operate on explicit sequences of flows. Instead, they apply one-dimensional convolutions over feature dimensions or short, fixed windows extracted from individual flows, learning \emph{spatio-temporal representations at the flow level}. These methods capture local temporal or statistical structure embedded in engineered features rather than long-range dependencies across flow sequences.

Several studies show that TCNs applied to aggregated or windowed flow features achieve competitive detection performance with substantially lower computational cost than sequence-level models~\citep{nazre2024tcn,lopes2023network}.~\citet{nazre2024tcn} proposed a TCN model with residual blocks and dilated convolutions for Edge-IIoTset, achieving 97\% accuracy while outperforming CNN-LSTM and CNN-GRU hybrids. ~\citet{li2025lightweight} developed TCNSE, integrating TCN with Squeeze-and-Excitation modules for lightweight IoT deployment, achieving 98.4\% accuracy on CIC-IDS-2018 with only 20 selected features. Hybrid architectures combining convolutional layers with LSTM or attention mechanisms similarly use TCNs primarily as feature extractors rather than sequence learners~\citep{he2023time,alashjaee2025deep}. While effective in practice, such models remain fundamentally per-flow classifiers and are more accurately characterized as spatio-temporal feature learning approaches.

Lightweight variants based on sliding-window aggregation of per-host or per-service statistics further illustrate this paradigm. When combined with feed-forward classifiers and causal normalization, these approaches achieve robust detection with constant per-flow computational cost, making them suitable for resource-constrained environments such as IoT networks~\citep{li2025lightweight,guo2025research}. Their temporal awareness arises from engineered aggregation rather than explicit sequence modeling.

\subsection{Transformers and Self-Attention}

Transformer architectures model temporal dependencies using self-attention, allowing each flow to attend to all others within a sequence. By removing recurrent dependencies, transformers capture long-range interactions more effectively than RNNs. Flow-based transformer models employ feature embeddings, positional encodings, and causal masking to preserve temporal order while enabling expressive sequence modeling~\citep{manocchio2023flowtransformer}. Extensions incorporating contextual signals such as n-gram frequency further enhance representation capacity~\citep{han2023network}. ~\citet{koukoulis2025self} proposed a self-supervised transformer-based contrastive learning approach that achieves strong performance without relying on labeled attack data.

However, transformers incur quadratic computational complexity with respect to sequence length, limiting scalability in high-throughput environments. Sparse and linear attention variants mitigate this cost but typically trade off representational fidelity. Moreover, strict attention masking and time-based dataset splits are essential to prevent future flows from influencing predictions, as even minor violations can introduce severe temporal leakage~\citep{corsini2021evaluation}. 

\subsection{Unsupervised and Self-Supervised Temporal Representation Learning}

Unsupervised and self-supervised methods reduce reliance on labeled attack data by learning temporal representations through reconstruction or contrastive objectives. Masked context reconstruction using transformer architectures improves robustness to label noise and anomalous contamination, yielding AUC gains exceeding 9\% on datasets such as UNSW-NB15 and CICIDS-WED~\citep{ruids2023}. Self-supervised contrastive learning further enables detection of zero-day attacks and adaptation to concept drift by maximizing agreement between augmented flow sequences~\citep{koukoulis2025self}. ~\citet{guerra2025graphids} proposed GraphIDS, combining graph representation learning and transformer-based autoencoder to capture both local and global structural patterns.

The effectiveness of these approaches critically depends on augmentation design and strict enforcement of temporal isolation, as inappropriate masking or sampling strategies can reintroduce information leakage and invalidate evaluation results. ~\citet{koukoulis2025self} demonstrated that self-supervised learning with carefully designed augmentations can significantly improve detection of unseen attacks.

\subsection{Mitigating Temporal Leakage}
\label{subsec:temporal_leakage}

Temporal leakage occurs when future information influences training or feature computation, producing overly optimistic performance estimates that fail to generalize to operational environments. Prior work identifies several essential mitigation strategies:

\begin{itemize}
    \item \textbf{Per-split normalization}: Normalization statistics must be computed independently for each data split using only past data~\citep{arp2022dos}.
    \item \textbf{Lagged statistics}: Running statistics should incorporate explicit temporal delays to prevent self-influence~\citep{corsini2021evaluation}.
    \item \textbf{Causal masking}: Attention mechanisms must strictly prohibit access to future flows, and positional encodings must encode only past context.
    \item \textbf{Time-based splits}: Chronological partitioning better reflects real-world deployment conditions than random splits~\citep{arp2022dos,cantone2024crossdataset}.
    \item \textbf{Cross-dataset evaluation}: Testing on temporally and distributionally distinct datasets provides a stronger assessment of generalization~\citep{cantone2024crossdataset}.
\end{itemize}

Even architectures that combine causal convolutions with masked attention remain vulnerable to leakage if evaluation protocols are flawed~\citep{li2025lightweight,guo2025research}. As emphasized by \citet{arp2022dos}, methodological rigor is a prerequisite for meaningful temporal modeling.

In summary, temporal modeling enables NIDS to capture sequential attack structure that is fundamentally inaccessible to per-flow classifiers. However, only models that combine explicit sequence reasoning with leakage-free evaluation protocols provide reliable and deployable detection performance in real-world settings.

% -------------------------------------------------------------
% GRAPH MODELING
% -------------------------------------------------------------
\section{Graph Modeling}
\label{sec:graph}

Graph modeling addresses the limitations of per-flow intrusion detection by explicitly representing relational structure in network traffic. Instead of treating flows as isolated samples, graph-based approaches encode communication patterns as graphs, where nodes correspond to network entities (e.g., hosts, flows, packets) and edges represent interactions between them~\citep{zhong2024survey}. This formulation allows detection models to use relationships and contextual information that per-flow classifiers cannot capture, making graph modeling especially effective for coordinated, multi-stage, and stealthy attacks.

The effectiveness of graph-based NIDS depends on how network entities are represented, how interactions are encoded, and how graph structure is updated over time. This section reviews major graph modeling paradigms, including homogeneous and heterogeneous graph neural networks (GNNs), federated and privacy-preserving learning, and self-supervised graph representation learning. We also discuss cross-cutting challenges that limit scalability and deployment in real-world environments. Table~\ref{tab:graph_char} provides a structured comparison of graph-based NIDS architectures, relating their message-passing mechanisms to the classes of coordinated attacks they detect and to the causality and construction constraints imposed by time-evolving network graphs.

\begin{table*}[t]
\centering
\footnotesize
\setlength{\tabcolsep}{4pt}
\caption{Characteristics of Graph Context Methods}
\label{tab:graph_char}
\begin{tabularx}{\textwidth}{@{}p{1.9cm} Y Y Y Y p{2.4cm}@{}}
\toprule
\textbf{Method} & \textbf{Key Mechanisms} & \textbf{Attack Profile} & \textbf{Key Limitations / Constraints} & \textbf{Critical Evaluation Requirements} & \textbf{Works} \\
\midrule
Homogeneous GNNs &
Message passing over single-type graphs (e.g., GraphSAGE, GAT). &
Scanning, botnets, lateral movement. &
Over-smoothing as depth increases; limited expressiveness for heterogeneous roles. &
Causal graph construction; exclude future edges and features. &
\citet{lo2022egraphsage,villegas2025dynamic,chang2021modified,zhang2025resacag} \\
\midrule
Heterogeneous GNNs &
Typed message passing over entity and event types. &
Cross-protocol and multi-stage attacks. &
Schema rigidity; modality imbalance; memory overhead. &
Consistent typing; aligned timestamps; schema-aware evaluation. &
\citet{schlichtkrull2018rgcn,farrukh2024xgnid,efficientgnn2023} \\
\midrule
Dynamic GNNs &
Time-evolving graphs with temporal message passing. &
Evolving botnets, transient C\&C. &
Snapshot instability; unbounded graph growth without pruning. &
Exclude future neighbors; time-ordered splits. &
\citet{mir2025dynkdd,duan2023dynamic,messai2025continuum} \\
\midrule
Federated GNNs &
Distributed training over local subgraphs. &
Cross-domain detection under privacy constraints. &
Non-IID drift; communication overhead. &
Cross-silo evaluation; secure aggregation awareness. &
\citet{Tfaily2025fedgatsage,nguyen2024fedmse} \\
\midrule
Self-Supervised Graph &
Edge feature reconstruction; random node/edge permutation or masking; negative edge sampling; link prediction. &
Sparse or weakly labeled attacks. &
Risk of structural leakage; large unlabeled graphs required. &
Masking must not reveal full adjacency; cross-dataset testing. &
\citet{guerra2025graphids,xu2022self,langendonck2024pptgnn} \\
\bottomrule
\end{tabularx}
\end{table*}

\subsection{Homogeneous Graph Neural Networks}

Early graph-based NIDS approaches model network traffic as homogeneous graphs, where all nodes and edges share the same type. Typically, hosts or IP addresses are represented as nodes and flows as edges connecting communicating endpoints. Message-passing GNNs such as GraphSAGE aggregate neighborhood information to learn node embeddings for intrusion detection in enterprise and IoT networks~\citep{lo2022egraphsage,villegas2025dynamic}. Graph Attention Networks (GATs) further improve expressiveness by learning attention weights over neighbors, at the cost of increased computational overhead~\citep{mir2025dynkdd}.

Several works extend homogeneous GNNs with architectural refinements. E-GraphSAGE introduces an edge-centric formulation in which flow records are directly mapped to graph edges, improving botnet detection in resource-constrained IoT settings~\citep{lo2022egraphsage}. Other variants incorporate residual connections, adaptive sampling, or attention mechanisms to better capture non-linear dependencies and improve generalization~\citep{chang2021modified,zhang2025resacag}. ~\citet{villegas2025dynamic} proposed dynamic graph modeling for IoT intrusion detection using GNNs, achieving strong performance on evolving network topologies.~\citet{geafl2025} introduced GEAFL-IDS, a graph edge attention approach with focal loss that addresses class imbalance in intrusion detection datasets.

Self-supervised learning has also been explored in homogeneous graph settings. Methods such as GraphIDS and Anomal-E employ edge features reconstruction and negative graph (corrupted graph) embedding discrimination objectives to reduce reliance on labeled data and improve robustness to noisy annotations~\citep{guerra2025graphids,xu2022self}. While effective, these approaches require large unlabeled graphs and careful masking strategies to prevent structural information leakage.

\subsection{Heterogeneous Graph Neural Networks}

Heterogeneous GNNs extend graph-based NIDS by modeling multiple node and edge types with relation-specific parameters, enabling richer representation of network semantics. Relational Graph Convolutional Networks (R-GCNs) provide a foundational framework for this paradigm by supporting typed message passing across diverse relations~\citep{schlichtkrull2018rgcn,langendonck2024pptgnn}.

Recent NIDS implementations exploit heterogeneity to integrate multiple data modalities. Some approaches construct graphs that combine dynamic traffic features with static vulnerability information, using heterogeneous GraphSAGE architectures followed by lightweight classifiers to achieve low-latency inference~\citep{efficientgnn2023}. XG-NID further extends this idea by modeling flows and packets as distinct node types connected via encapsulation edges, preserving payload-level context alongside flow statistics~\citep{farrukh2024xgnid}. Although effective, such designs introduce substantial memory overhead and training instability due to modality imbalance. ~\citet{jiang2025hgnn} provided a comprehensive survey of heterogeneous GNNs for cybersecurity anomaly detection, identifying key challenges in standardization and reproducibility.

Despite their expressive power, heterogeneous GNN-based NIDS lack standardization. Surveys highlight significant variability in graph construction strategies, node and edge typing, and evaluation protocols, complicating reproducibility and cross-paper comparison~\citep{bilot2023survey,jiang2025hgnn}. Moreover, the computational cost of heterogeneous message passing poses challenges for deployment in high-speed networks.

\subsection{Federated and Privacy-Preserving Graph Learning}

Federated graph learning addresses privacy and data governance constraints by enabling collaborative model training across distributed network domains without sharing raw traffic data. In this setting, each participant trains a local GNN on its subgraph and periodically shares model updates with a central aggregator. FedGATSage exemplifies this approach by combining federated learning with a heterogeneous GNN architecture for IoT intrusion detection, using community-aware abstraction to reduce communication overhead~\citep{Tfaily2025fedgatsage}. However, restricting global context in this manner degrades detection performance for cross-community attacks. ~\citet{nguyen2024fedmse} proposed FedMSE, a semi-supervised federated learning approach combining shrink autoencoders with centroid one-class classifiers for IoT NIDS.

Privacy guarantees can be strengthened through secure aggregation and differential privacy mechanisms, though these techniques introduce additional accuracy and convergence trade-offs~\citep{BouayadAMAB24}. Overall, federated graph learning provides a viable solution for privacy-sensitive environments, but remains constrained by communication costs, partial observability, and limited global awareness.

\subsection{Self-Supervised Graph Representation Learning}

Self-supervised graph learning aims to learn transferable representations from unlabeled traffic graphs, reducing dependence on costly labeled data. ~\citet{xu2022self} proposed Anomal-E, a self-supervised network intrusion detection system based on GNNs that learns effective representations without labeled attack data. It trains a graph encoder (E-GraphSAGE) using edge‑feature reconstruction and discrimination between real and corrupted graphs, encouraging embeddings that capture both local edge attributes and global structural patterns. For anomaly detection, they utilized four unsupervised algorithms: principal component analysis (PCA), isolation forest (IF), cluster-based local outlier factor (CBLOF), and histogram-based outlier score (HBOS). Each algorithm was trained on edge embeddings derived from the training graph, which served as the input features for learning normal behavioral patterns and identifying anomalies. ~\citep{guerra2025graphids} introduce GraphIDS, a self-supervised framework for network intrusion detection that jointly optimizes graph representation learning and anomaly detection through an end-to-end architecture combining an inductive graph neural network (E-GraphSAGE) with a Transformer-based masked autoencoder. Across diverse NetFlow benchmarks, GraphIDS acheive up to 99.98\% PR-AUC and 99.61\% macro F$_1$, exceeding baseline methods by 5--25 percentage points.

Despite these promising results, self-supervised graph learning for NIDS remains underexplored. Open challenges include the design of effective graph augmentations, prevention of temporal and structural leakage during pre-training, and robustness under cross-dataset evaluation~\citep{xu2022self,lin2024egracl}. Addressing these issues is critical for deploying self-supervised GNNs in dynamic operational environments. ~\citet{langendonck2024pptgnn} introduced PPT-GNN, a practical pre-trained spatio-temporal GNN that enables near real-time predictions and better cross-network generalization.  It represents NetFlow traffic as heterogeneous graphs built from short time-based sliding windows, enriched with intra- and inter-window temporal edges. Then it uses a self-supervised link-prediction task for every edge in the graph.

\subsection{Cross-Cutting Challenges}

Several fundamental challenges hinder the deployment of graph-based NIDS in practice. Graph construction remains highly dataset-specific, with temporal consistency being particularly critical: communication graphs must preserve chronological flow ordering to prevent leakage from future observations, requiring causal neighbor sampling and time-aware edge formation~\citep{arp2022dos}. 

Scalability is another major constraint, as operational networks generate traffic volumes that overwhelm full-graph training paradigms. Sampling, compression, and distributed training alleviate this burden but introduce trade-offs in detection fidelity~\citep{Tfaily2025fedgatsage}. ~\citet{zhong2023dynamic} proposed dynamic multi-scale topological representation learning to address scalability in network intrusion detection.

Finally, cross-dataset evaluation reveals severe generalization limitations. Models trained on benchmarks such as CIC-IDS-2017 often perform poorly when transferred to datasets like CTU-13, exposing strong dataset-specific biases in learned graph structures~\citep{cantone2024crossdataset,sharafaldin2018toward,garcia2014ctu13}. Bridging this gap remains an open research problem.~\citet{messai2025continuum} proposed CONTINUUM, a spatio-temporal GNN framework for APT attack detection that addresses both temporal and relational dependencies.

In summary, graph modeling provides a powerful framework for capturing relational and contextual information in network intrusion detection. Homogeneous and heterogeneous GNNs, along with federated and self-supervised extensions, expand detection capability under complex and constrained settings. Realizing their full potential requires (i) principled, time-consistent graph construction, (ii) scalable learning and summarization under streaming traffic, and (iii) standardized evaluation protocols that demonstrate robustness across datasets and deployment conditions.

% -------------------------------------------------------------
% MULTIMODAL AND MULTI-RESOLUTION
% -------------------------------------------------------------
\section{Multimodal and Multi-Resolution Architectures}
\label{sec:multimodal}

Modern networks emit many partially overlapping signals: flow summaries, packet content (when available),
control-plane metadata (DNS, TLS, routing), and host-side logs. Architectures that rely on a single signal
or a single granularity often miss attacks whose evidence is \emph{distributed} across sources or \emph{split}
across time scales~\citep{farrukh2024xgnid,messai2025continuum}. Two complementary design patterns address this gap:

\textbf{Multimodal fusion} combines \emph{different telemetry sources} (``more sensors'') to reconstruct richer context.
\textbf{Multi-resolution modeling} combines \emph{different granularities of activity} (``more zoom levels'') such as
packet $\rightarrow$ flow $\rightarrow$ session $\rightarrow$ host, so that the detector can reason about both fine-grained
and coarse-grained behaviors. These patterns are orthogonal and are often combined in a single system; we separate them
here to make the trade-offs explicit.

\subsection{Multimodal Fusion}
\label{sec:multimodal_fusion}

Multimodal NIDS ingest heterogeneous streams and learn a joint representation that is more informative than any single
modality. In practice, the key question is \emph{what gets fused and when}: early fusion (concatenating raw features), mid-level fusion (aligning learned embeddings), or late fusion (ensembling modality-specific detectors). Each choice trades off robustness, interpretability, and deployment cost. Table~\ref{tab:multimodal_char} categorizes multimodal fusion strategies in NIDS, linking each fusion mechanism to the attack patterns it exposes and to the alignment, privacy, and missing-modality constraints that must be respected during evaluation.

\begin{table*}[t]
\centering
\footnotesize
\setlength{\tabcolsep}{4pt}
\caption{Characteristics of Multimodal Fusion Methods}
\label{tab:multimodal_char}
\begin{tabularx}{\textwidth}{@{}p{2cm} Y Y Y Y p{2.4cm}@{}}
\toprule
\textbf{Fusion / Mechanism} & \textbf{Key Mechanisms} & \textbf{Attack Profile} & \textbf{Key Limitations} & \textbf{Evaluation Requirements} & \textbf{Works} \\
\midrule
Flow + Packet &
CNN/graph fusion of payload (when available) and flow stats. &
Encrypted malware via handshake side-channels; payload signatures only with DPI. &
Encryption and privacy constraints; packet storage cost. &
Explicit modality availability; causal alignment. &
\citet{farrukh2024xgnid,shi2023mhpnnids,kiflay2024multimodal,xu2025fewshot} \\
\midrule
Flow + DNS/TLS &
Control-plane fusion via attention or graphs. &
DGA malware, encrypted C\&C. &
Clock skew; missing modalities. &
Missing-modality-aware evaluation. &
\citet{xu2025fewshot,zha2025dmids,zhang2025norns} \\
\midrule
Flow + Host Logs &
Late fusion across domains. &
Host compromise with exfiltration. &
Privacy concerns; timestamp alignment. &
Prevent future-log leakage. &
\citet{toniots,wali2023meta} \\
\midrule
Flow + Alerts &
Alert-context graphs. &
Multi-stage campaigns. &
Error propagation from upstream IDS. &
Strict causal windows. &
\citet{messai2025continuum,govindarajan2025} \\
\bottomrule
\end{tabularx}
\end{table*}

\paragraph{Flow--packet fusion.}
The most common setting fuses flow statistics with packet-level content. XG-NID, for example, treats flows and packets
as distinct node types in a heterogeneous graph and jointly models temporal flow features with payload-derived
representations~\citep{farrukh2024xgnid}. Related hybrid architectures (e.g., CNN--LSTM ensembles) similarly combine
tabular flow features with byte/packet representations and consistently improve benchmark accuracy over single-modality
baselines~\citep{shi2023mhpnnids,kiflay2024multimodal}. Recent designs further target efficiency: transformer-style
models and linear state-space variants integrate packet sequences while reducing quadratic attention costs, making
pretraining and inference more practical on commodity hardware~\citep{zhang2025norns}. ~\citet{xu2025fewshot} proposed a multimodal fusion based few-shot network intrusion detection system combining traffic feature graphs and network feature sets.

\paragraph{Control-plane and host-side signals.}
Payload is often unavailable due to encryption and privacy constraints, so many deployments instead fuse control-plane
signals (DNS requests, TLS handshakes, BGP updates) and host telemetry (system logs, authentication events).
Few-shot and robust multimodal frameworks show that adding these signals can substantially improve detection in
complex scenarios where flow statistics alone are ambiguous~\citep{xu2025fewshot,zha2025dmids}.~\citet{zhang2025norns} proposed Norns, leveraging multi-modal Mamba for efficient network intrusion detection by integrating comprehensive flow features and packet bytes.

\paragraph{Practical bottlenecks.}
Multimodal fusion is only as good as its alignment. Streams arrive with different clocks and sampling rates, and even
small misalignment can cause attention modules to focus on the wrong evidence. As a result, effective fusion typically
requires explicit temporal synchronization and mechanisms that prevent ``easy'' modalities (high-rate, high-signal) from dominating ``rare'' modalities (low-rate but high-value). These challenges often drive design choices toward late fusion or lightweight alignment layers in real deployments.

\subsection{Multi-Resolution Modeling}
\label{sec:multiresolution}

Multi-resolution architectures keep the underlying signal(s) fixed but aggregate activity at multiple granularities. The motivation is straightforward: coarse views highlight volumetric anomalies and persistent shifts, while fine views preserve short-lived indicators and low-and-slow behavior. A detector that only sees one scale must implicitly learn the other scales, which is difficult under concept drift. Table~\ref{tab:multires_char} summarizes multi-resolution architectures that aggregate information across packets, flows, sessions, and hosts, and details how their hierarchical design affects both attack detectability and causality-preserving evaluation.

\begin{table*}[t]
\centering
\footnotesize
\setlength{\tabcolsep}{4pt}
\caption{Characteristics of Multi-Resolution Methods}
\label{tab:multires_char}
\begin{tabularx}{\textwidth}{@{}p{2.1cm} Y Y Y Y p{2.4cm}@{}}
\toprule
\textbf{Hierarchy} & \textbf{Key Mechanisms} & \textbf{Attack Profile} & \textbf{Key Limitations} & \textbf{Evaluation Requirements} & \textbf{Works} \\
\midrule
Packet$\to$Flow &
Hierarchical aggregation from bytes to flows. &
Early payload anomalies. &
Requires packet visibility; memory growth. &
Causal aggregation only. &
\citet{wu2025uninet,hasan2025iot,he2019novel} \\
\midrule
Flow$\to$Host &
Per-host sliding windows. &
Scanning, volumetric abuse. &
State growth with host count. &
Lagged windows only. &
\citet{messai2025continuum,zhong2023dynamic} \\
\midrule
Session-Level &
Flow grouping with sequence models. &
Application-layer attacks. &
Session boundary ambiguity. &
Avoid future-dependent labels. &
\citet{li2025lightweight,guo2025research} \\
\midrule
Hierarchical Pooling &
Graph pooling across scales. &
Multi-scale coordinated attacks. &
Transductive leakage risk. &
Inductive pooling only. &
\citet{shi2023mhpnnids,hasan2025iot} \\
\bottomrule
\end{tabularx}
\end{table*}

\paragraph{Hierarchies over packet/flow/session/host.}
Multi-resolution systems build hierarchical summaries and let the model decide which scale matters for each decision. UniNet is an example of this direction: it integrates multi-granular representations with attention to generalize
across diverse attack types~\citep{wu2025uninet}. In IoT settings, lightweight designs combine tabular summaries with image-like encodings to preserve real-time throughput while still capturing complementary patterns across granularities~\citep{hasan2025iot}. \citet{hasan2025iot} proposed a cost-sensitive multimodal approach for generalizable IoT intrusion detection.

\paragraph{Multi-resolution graphs and evolving topology.}
When the network structure itself is part of the signal, multi-resolution modeling often becomes graph-based:
edges capture interactions, and hierarchical pooling or temporal modules track changes in topology.
Spatio-temporal graph architectures can detect coordinated, multi-stage activity by aggregating evidence across
both space (who talks to whom) and time (when and in what order)~\citep{messai2025continuum}. The main technical
challenge is \emph{resolution fusion}: naive concatenation lets high-volume resolutions dominate, while hierarchical
pooling and attention gating are needed to preserve rare but critical patterns from underrepresented scales~\citep{he2019novel,shi2023mhpnnids}. ~\citet{messai2025continuum} demonstrated this through CONTINUUM, which uses spatio-temporal graph neural networks for APT attack detection.

\subsection{Cross-Cutting Challenges}
\label{sec:multimodal_challenges}

Three issues repeatedly limit operational deployment.

\textbf{Benchmarks and reproducibility.}  Public datasets rarely provide synchronized multimodal telemetry across
multiple granularities, forcing reliance on proprietary collections and making results hard to reproduce~\citep{wali2023meta,zhang2019multilayer}.

\textbf{Cost and privacy.}  Multimodal and multi-resolution models can be expensive to run on high-speed links.
Aggressive sampling, compression, or selective logging helps throughput but risks removing the very evidence these
models were designed to capture. Meanwhile, privacy regulations often restrict sensitive modalities (payloads, host logs),
motivating privacy-preserving or federated solutions that introduce communication and utility trade-offs.

\textbf{Generalization under drift.}  Models trained on controlled benchmarks often degrade under concept drift and
novel attack patterns. Real progress therefore requires evaluation protocols that test across environments and enforce
temporal integrity, so that ``context'' does not become a source of leakage rather than robustness.

% -------------------------------------------------------------
% DATASETS AND EVALUATION
% -------------------------------------------------------------
\section{Datasets and Evaluation Methodology}
\label{sec:evaluation}

Reliable evaluation of context-aware network intrusion detection systems (NIDS) requires representative datasets, causality-preserving protocols, and metrics that reflect both detection effectiveness and operational constraints. Unlike conventional classification tasks, NIDS evaluation must account for temporal dependencies, severe class imbalance, adversarial behavior, and non-stationary traffic distributions. This section reviews the limitations of existing datasets and summarizes best practices for robust and realistic evaluation.

\subsection{Dataset Limitations}

Most widely used NIDS benchmarks exhibit methodological flaws that compromise evaluation validity. CIC-IDS-2017 and CSE-CIC-IDS-2018, for example, contain labeling errors and unrealistic temporal separation between attack and benign traffic~\citep{Engelen20217,lanvin2022errors}. A systematic analysis of seven popular benchmarks identifies six recurring design issues: limited feature diversity, high feature interdependence, ambiguous ground truth, collapse of traffic distributions, artificial diversity generation, and labeling inaccuracies~\citep{bad_smells}. These flaws correlate strongly with poor generalization in operational environments. Most benchmarks rely on synthetic or testbed traffic with limited attack diversity. While real-world datasets such as MAWIFlow provide high-fidelity traffic with natural concept drift~\citep{schraven2025mawiflow}, they lack comprehensive ground truth due to privacy constraints. IoT-specific datasets capture domain-specific traffic patterns but transfer poorly to enterprise settings~\citep{Hawawreh2022xiiotid}. Among recent efforts, ToN-IoT offers a multimodal testbed combining IoT telemetry, host logs, and NetFlow traces~\citep{toniots}, but introduces challenges in temporal alignment and cross-dataset comparability.

\citet{schraven2025mawiflow} introduced MAWIFlow, a flow-based benchmark derived from MAWILAB v1.1 that enables realistic evaluation with temporally distinct samples spanning 2011--2021. ~\citep{ghosh2025tacnet} evaluated TACNet on multiple datasets including CIC IoT-DIAD 2024 and TabularIoTAttack-2024, demonstrating cross-dataset robustness.

Dataset diversity plays a central role in model generalization and can be quantified along three dimensions: intra-class variation, inter-class separation, and domain-shift robustness. Diversity-aware metrics such as the Vendi score and Jensen--Shannon divergence provide complementary measures that correlate with improved transferability~\citep{nougnanke2025diversity,friedman2023vendi}. These measures, combined with classification complexity and dataset cartography techniques~\citep{lorena2019complexity,swayamdipta2020dataset}, offer a principled framework for assessing benchmark difficulty beyond raw accuracy.

\subsection{Evaluation Rigor and Robustness}

Evaluation protocols must prevent information leakage and reflect deployment conditions. Best practices include:

\begin{itemize}
    \item \textbf{Chronological splitting}: partition data in temporal order to preserve causality and capture distribution shifts~\citep{arp2022dos,corsini2021evaluation}.
    \item \textbf{Per-split normalization}: compute normalization statistics exclusively on training data and apply them separately to validation and test sets~\citep{arp2022dos}.
    \item \textbf{Causal modeling}: enforce lagged statistics for feature engineering and causal masking in sequence and graph models~\citep{corsini2021evaluation}.
    \item \textbf{Cross-dataset validation}: evaluate models across heterogeneous datasets to identify dataset-specific biases~\citep{cantone2024crossdataset}.
    \item \textbf{Comprehensive metrics}: report macro-F1, AUROC, precision and recall at $K$, false alarm rate, and per-attack performance in addition to accuracy~\citep{binbusayyis2020comprehensive}.
\end{itemize}

Beyond static evaluation, NIDS operate in adversarial and non-stationary environments. Gradient-based evasion attacks can bypass ML-based detectors by perturbing features within constrained budgets~\citep{mohammadian2023gradient,apruzzese2024adversarial}. At the same time, concept drift driven by software updates, evolving user behavior, and network reconfiguration progressively degrades detection performance. Defense strategies combine adversarial training with ensemble learning and preprocessing techniques such as denoising autoencoders~\citep{awad2025enhanced}. Adaptive systems like INSOMNIA integrate active learning and explainable AI for continuous model updating~\citep{andresini2021insomnia}. \citet{shyaa2025reinforcement} proposed reinforcement learning-based voting for feature drift-aware intrusion detection using incremental learning frameworks.

Rigorous evaluation of such systems must therefore include both adversarial stress testing and longitudinal analysis on non-stationary data. Formal verification can provide additional robustness guarantees in restricted settings~\citep{flood2025formal}, but remains computationally expensive for large-scale deployments.~\citet{mehta2025meda} proposed MEDA, a Mixture-of-Experts based concept drift adaptation framework for in-vehicle network intrusion detection.

Overall, current evaluation practices are fundamentally constrained by the scarcity of public, diversified, and well-documented datasets that reflect real-world adversarial dynamics and concept drift. Progress in context-aware NIDS requires coordinated efforts toward representative benchmark construction, diversity-aware evaluation metrics, and standardized protocols that enforce temporal causality and prevent information leakage~\citep{arp2022dos}. Only under such rigorous evaluation frameworks can research advances translate into reliable and generalizable intrusion detection systems.

% -------------------------------------------------------------
% IMPLEMENTATION AND DEPLOYMENT
% -------------------------------------------------------------
\section{Implementation and Deployment Considerations}
\label{sec:implementation}

Deploying context-aware NIDS in operational environments requires balancing detection performance with systems-level constraints, including computational efficiency, memory usage, robustness, and adaptability to evolving traffic. Unlike offline experimental settings, production deployments must operate under strict latency budgets, limited resources, and continuous data streams. This section synthesizes key implementation considerations that bridge the gap between research prototypes and real-world systems.

\subsection{Streaming State Management}

Temporal and graph-based models maintain state across flows through hidden states in recurrent networks, context windows in transformers, or neighborhood caches in GNNs. In streaming environments, this state must be managed within bounded memory while preserving detection capability. 

Common strategies include sliding windows over recent flows, which trade off memory footprint against temporal coverage, and stateful caching of per-host aggregates with time-to-live semantics to ensure automatic expiration of stale information. Graph-based systems further rely on dynamic neighborhood management, using sampling or pruning to bound graph size. Sketch-based and approximate neighbor representations offer promising directions, but their impact on detection fidelity remains underexplored in NIDS settings~\citep{corsini2021evaluation}. Lag buffers are often required to compute delayed statistics for normalization and feature engineering, ensuring temporal causality and preventing self-normalization bias~\citep{corsini2021evaluation}. ~\citet{meganathan2025adwise} proposed ADWISE, an adaptive drift-aware windowing intrusion detection system with optimization for handling concept drift in streaming data.

\subsection{Computational Efficiency}

Operational NIDS must satisfy strict latency and throughput constraints, particularly in edge or high-speed network deployments. Real-time systems typically require inference latency below a few milliseconds per flow, making computational efficiency a first-order design objective.

Model compression techniques such as quantization and pruning significantly reduce inference cost with limited accuracy degradation. Quantization-aware training and neural architecture search further optimize models for specific latency and energy targets~\citep{umar2025energy,chitty2022neural,BouayadAIB24}. Architectural choices also have a major impact: GraphSAGE provides faster inference than attention-based GNNs, while temporal convolutional networks and linear transformers offer favorable accuracy--latency trade-offs for sequence modeling~\citep{Kurdianto2025}. Hardware acceleration using GPUs, FPGAs, or ASICs improves throughput and energy efficiency, complemented by energy-aware scheduling and workload partitioning~\citep{Abdulganiyu2023,Ahmad2020}.~\citet{hu2025dwoids} proposed DWOIDS, an online intrusion detection system based on dual adaptive windows and Hoeffding tree classifiers for concept drift adaptation.

\subsection{Adaptive Learning and Concept Drift}

Operational NIDS must cope with label scarcity and non-stationary traffic distributions. Self-supervised learning reduces reliance on labeled data by leveraging abundant unlabeled traffic through masked prediction and contrastive objectives, enabling robust representation learning without relying on labeled data~\citep{guerra2025graphids,koukoulis2025self}. 

Continual learning techniques address concept drift by enabling incremental adaptation without catastrophic forgetting. Approaches include replay buffers, regularization-based updates, and periodic fine-tuning~\citep{nakip2023online}. Adaptive frameworks such as INSOMNIA integrate active learning and explainable AI to support continuous model refinement with analyst feedback~\citep{andresini2021insomnia}. Generative approaches such as GenCoder++ further use variational autoencoders to synthesize new samples, improving robustness to evolving attack patterns~\citep{smolin2025gencoder++}. ~\citet{hu2025dwoids} demonstrated that dual adaptive windows with momentum-based drift detection can significantly improve adaptation to concept drift.

\subsection{Data Augmentation and Traffic Generation}

Synthetic data generation mitigates class imbalance and enables stress testing under controlled conditions. Frameworks such as ConCap generate flow traces with configurable attack mixtures, supporting systematic evaluation of detection robustness~\citep{concap2025}. Generative adversarial networks and other augmentation techniques increase training diversity and improve detection of rare attack patterns~\citep{park2023gan,muller2025augmentation}. 

While synthetic data cannot replace real traffic, combining generation with domain adaptation techniques helps bridge the gap between controlled training environments and operational networks. ~\citet{agrate2024adaptive} proposed adaptive ensemble learning for intrusion detection systems under concept drift.

\subsection{Security and Privacy Preservation}

ML-based NIDS are vulnerable to adversarial manipulation of input features, which can significantly degrade detection performance~\citep{mohammadian2023gradient,apruzzese2024adversarial}. Defense mechanisms include adversarial training, ensemble methods, invariant feature learning, and certified robustness through formal verification, although scalability remains a challenge~\citep{flood2025formal,sauka2022adversarial}. \citet{electronics2025advnn} proposed ADV\_NN, an adversarially trained neural network that maintains over 80\% accuracy under PGD and FGSM attacks.

Privacy constraints further complicate deployment, especially in multi-organizational environments. Privacy-preserving frameworks such as PriMIA use homomorphic encryption to protect both traffic data and model parameters~\citep{primia2024}. Federated learning with secure aggregation enables collaborative training across sites without raw data sharing, but introduces communication overhead and reduced global context~\citep{sorour2025federated,li2023federated,idrissi2023fed,BouayadAMAB24}. ~\citet{khan2025xai} provided a systematic review on explainable AI-based intrusion detection systems for Industry 5.0 and adversarial XAI.

\subsection{Operational Integration}

Effective deployment requires seamless integration with existing security infrastructure and human workflows. Zero-trust architectures demand continuous authentication and authorization, necessitating tight coupling between NIDS and identity management systems for context-aware alert prioritization~\citep{nist2019zero_trust}. 

Explainable AI frameworks generate interpretable insights through local and global explanations, reducing analyst cognitive load and supporting trust in automated decisions~\citep{alabbadi2025xaiids}. Human-in-the-loop systems combine automated detection with analyst feedback, enabling adaptation to novel threats and reducing false positives. However, standardized interpretability metrics and systematic evaluation of performance--explainability trade-offs remain open research problems. ~\citet{khan2025xai} surveyed XAI methods for enhancing IDS, highlighting the need for transparent decision-making in security operations.

Overall, successful deployment of context-aware NIDS requires co-design of algorithms, systems, and operational workflows. Detection models must be engineered not only for accuracy, but also for scalability, robustness, privacy, and human usability. Without addressing these systems-level constraints, advances in context-aware modeling are unlikely to translate into reliable and sustainable security solutions.

% -------------------------------------------------------------
% CONCLUSION
% -------------------------------------------------------------
\section{Conclusion and Future Directions}
\label{sec:conclusion}

Context-aware deep learning provides a principled direction for network intrusion detection systems that operate on flow telemetry without payload inspection. By modeling temporal, relational, multimodal, and multi-resolution dependencies, these approaches capture attack patterns that are fundamentally inaccessible to per-flow classifiers. Across the literature, context-aware architectures consistently demonstrate improved detection of multi-stage, distributed, and stealthy threats.

However, empirical progress remains constrained by methodological and practical limitations. First, evaluation protocols must rigorously enforce temporal causality and prevent information leakage. Numerous reported gains disappear under realistic settings with chronological splits, per-split normalization, and causal feature construction, highlighting the fragility of current benchmarking practices. Second, dataset quality remains a dominant bottleneck. Label noise, synthetic biases, limited diversity, and poorly documented collection procedures compromise both training and evaluation. Diversity-aware metrics such as the Vendi score and dataset cartography provide useful diagnostic tools, but cannot substitute for representative, well-curated benchmarks that reflect real-world traffic conditions.

Third, cross-dataset generalization remains unresolved. Most context-aware models exhibit strong performance only within the distribution of their training data and fail to transfer reliably across datasets~\citep{AlamiIMBYB23}. Techniques such as heterogeneity-aware validation, domain adaptation, and federated learning offer promising directions, but lack systematic evaluation at scale. Without reliable generalization, improvements on individual benchmarks provide limited operational value.

Fourth, deployment feasibility introduces additional constraints that are rarely addressed in experimental studies. Context-aware NIDS must balance accuracy with latency, memory usage, energy consumption, and privacy requirements. Lightweight architectures, quantization, streaming state management, and self-supervised learning reduce resource demands, but adversarial robustness and long-term adaptation to concept drift remain underexplored. Integrating adversarial defenses, invariant representation learning, and continual learning is essential for sustaining detection performance in dynamic environments.

Looking forward, meaningful progress in context-aware NIDS requires a shift from isolated model development toward community-driven infrastructure. Priority research directions include: (i) the release of public, synchronized multimodal datasets with quantified diversity and realistic temporal structure; (ii) standardized evaluation protocols that enforce causality, prevent leakage, and mandate cross-dataset testing; (iii) open-source pipelines for reproducible benchmarking; and (iv) systematic study of deployment constraints under adversarial and non-stationary conditions. Bridging the gap between algorithmic advances and operational realities is critical for translating context-aware learning into reliable, real-world intrusion detection systems~\citep{aldweesh2020deep,wu2024current,lansky2023deep,aleesa2019review,liao2023survey}.

\bibliographystyle{plainnat}
\bibliography{references}
\end{document}